\documentstyle[emulateapj]{article}

\begin{document}

\title{THE MINOR-MERGER DRIVEN NUCLEAR ACTIVITY IN SEYFERT GALAXIES: 
       A STEP TOWARD THE SIMPLE UNIFIED FORMATION MECHANISM OF 
       ACTIVE GALACTIC NUCLEI IN THE LOCAL UNIVERSE}

\author{Yoshiaki Taniguchi}

\vspace {1cm}

\affil{Astronomical Institute, Graduate School of Science, 
       Tohoku University, Aoba, Sendai 980-8578, Japan}


\begin{abstract}
Seyfert nuclei are typical active galactic nuclei (AGNs) in the local
universe, which  are thought to be powered by gas accretion onto a
supermassive black hole (SMBH). 
The dynamical effect exerted either by non-axisymmetric gravitational potential
(such as a stellar bar) or by interaction with other galaxies has been often 
considered to cause the efficient gas fueling. However, recent systematic 
studies for large samples of Seyfert nuclei have shown that;
1) Seyfert nuclei do not prefer barred galaxies as their hosts,
and 2) only $\simeq$ 10\% of Seyfert galaxies have companion galaxies. 
These findings raise again a question; ^^ ^^ What is important 
in the fueling of nuclear activity ?"
Here we suggest that an alternative fueling mechanism, the minor-merger
(i.e., mergers between a gas-rich disk galaxy and a bound, less-massive
satellite galaxy) driven fueling appears consistent with 
almost all important observational properties of Seyfert galaxies.
Nucleated (i.e., either a SMBH or a dense nuclear star cluster) galaxies and
satellites seem necessary to ensure that the gas in the host disk is surely
fueled into the very inner region (e.g., $\ll$ 1 pc).
Taking account that local quasars may be formed by major mergers between/among
galaxies, we propose that all the AGNs in the local universe
were made either by minor or by major mergers.
\end{abstract}




\section{INTRODUCTION}

Seyfert galaxies are typical examples with active galactic nuclei (AGNs) in the 
local universe and thus they have been well studied from various observational
points of view. It is known that Seyfert nuclei are associated with disk galaxies.
If the majority of disk galaxies have a SMBH in their nuclei
(e.g., Kormendy et al. 1998 and references therein),
the most important problem is how the gas can be efficiently fueled onto the SMBH
because the central engine of AGNs is thought to be powered by the accretion of gas
onto a SMBH (Rees 1984).
The dynamical effect exerted either by non-axisymmetric gravitational potential
such as a stellar bar or by interaction with other galaxies has been often 
considered to cause the efficient gas fueling (e.g., Noguchi 1988; 
Shlosman, Frank, \& Begelman 1989; Barnes \& Hernquist 1992;
Shlosman \& Noguchi 1993; see for a review Shlosman, Begelman, \& Frank 1990).
However, recent systematic studies for large samples of Seyfert nuclei 
have shown that the Seyfert galaxies do not always have either the bar 
structure or companion galaxies
(Mulchaey \& Regan 1997; Ho, Filippenko, \& Sargent 1997; Rafanelli et al. 1995;
De Robertis, Yee, \& Hayhoe 1998b).
Then we have not yet fully understood what causes the efficient gas fueling
onto a SMBH in the Seyfert nuclei. Here,
adopting an alternative fueling mechanism, the minor-merger driven
fueling (Roos 1981, 1985a, 1985b; Gaskell 1985; Hernquist 1989; 
Mihos \& Hernquist 1994; Hernquist \& Mihos 1995;
De Robertis et al. 1998b; Taniguchi 1997),
we examine whether or not this mechanism can explain all important observational
properties of Seyfert nuclei.

\section{A BRIEF REVIEW OF POSSIBLE FUELING MECHANISMS}

\subsection{Galaxy-Interaction Driven Gas Fueling}

The nuclear activity in interacting galaxies has been 
investigated in these two decades (Petrosyan 1982; Kennicutt \& Keel 1984; 
Keel et al. 1985; Dahari 1985; Cutri \& McAlary 1985; Bushouse 1986, 1987;
Telesco, Wolstencroft, \& Done 1988;  Smith \& Hintzen 1991;
Sekiguchi et al. 1992; Bergvall \& Johansson 1995).
In particular, the environmental properties of Seyfert galaxies have been 
one of key issues (Dahari 1984; Fuentes-Williams \& Stocke 1988;
MacKenty 1989; Laurikainen et al. 1994; Laurikainen \& Salo 1995;
Rafanelli et al. 1995; De Robertis, Hayhoe, \& Yee 1998a; De Robertis
et al. 1998b). Although early investigations suggested a possible 
excess of companion galaxies in Seyfert galaxies (e.g., Dahari 1984), 
this excess has not been confirmed (e.g., De Robertis et al. 1998b).
Although S2s still tend to have more companions with respect to normal 
disk galaxies, its statistical significance is $\simeq$ 95\% at most
(De Robertis et al. 1998b).

An important point noted here is that  {\it only} 10\% 
of the Seyfert galaxies have companion galaxies
(e.g., Rafanelli et al.  1995).  This means that the remaining $\sim$ 90\% of
the Seyfert galaxies have no comparable companion galaxy
and thus their nuclear activity is not related to galaxy interactions.
In addition, Keel (1996) found that  there is no preferred kind of interaction
(prograde, polar, or retrograde) among the Seyfert galaxies with physical
companions although the efficient fueling would occur in prograde interacting systems.
Hence, even if the nuclear activity in some Seyfert galaxies were triggered
by galaxy interactions, 
the tidal triggering appears to be a very minor mechanism.

\subsection{Bar Driven Gas Fueling}

The non-axisymmetric gravitational potential such as a stellar bar in a galactic disk
is considered to drive the mass transfer of interstellar medium
from the disk to the central region (Schwartz 1981; Norman 1990; Shlosman,
Frank, \& Begelman 1989; Wada \& Habe 1992, 1995). 
However, from an observational point of view, 
the excess of barred galaxies in Seyfert galaxies has been controversial
(Adams 1977; Simkin et al. 1980; Arsenault 1989; Moles, M\'arquez, \& P\'erez 1995;
Maiolino, Risaliti, \& Salvati 1998). 
Recently, Mulchaey \& Regan (1997) made a near-infrared imaging survey of samples
of Seyfert and normal galaxies and found that the incidence of bars in both 
the samples is quite similar; i.e., $\sim$ 70\% (see also Hunt et al. 1999). 
This means that Seyfert nuclei do not prefer
barred galaxies as their hosts. Another very important work was made by 
Ho, Filippenko, \& Sargent (1997) who analyzed optical spectra of more than 300
spiral galaxies in the nearby universe. 
They found that AGNs (Seyferts and LINERs\footnote{LINER = Low Ionization Nuclear
Emission-line Region (Heckman 1980).}) do not show 
any significant preference of barred galaxies. 
They also found that bars have a negligible effect on the strength of AGN.
In summary, the recent statistical studies based on the larger samples
have suggested that the bar-driven gas inflow is not a dominant mechanism for
triggering activity in Seyfert nuclei\footnote{If the dynamical lifetime of
bar structures is comparable to or shorter than the lifetime of AGNs, 
one finds no significant correlation between the presence of bars and AGNs.
However, one would prove that the efficient gas fueling into a central
$\ll$ 1 pc region is surely driven by the dynamical effect of bars
(see also footnote 3).}.

Very recently, Maiolino, Risaliti, \& Salvati (1998) presented evidence for
a strong correlation between the gaseous absorbing column density
towards S2  nuclei and the presence of a stellar bar in their host galaxies.
They also showed that strongly barred S2s
have an average $N_{\rm HI}$ (H {\sc i} column density) that is two orders of
magnitude higher than non-barred S2s.
Although these properties are quite interesting,
the main point stressed here is that a large-scale bar structure
has no relation to the gas fueling in the Seyfert nuclei
from the statistical studies described before. We also note that
the so-called bars-within-bars\footnote{Secondary (i.e., inner) bars and triaxial
bulges are found in some barred galaxies. These double-bar structures
have been sometimes discussed in relation to the bars-within-bars
mechanism proposed by Shlosman et al. (1989) (e.g., Friedli 1996). However,
such large-scale double bars may be more intimately related to
the stellar dynamical resonance in disk galaxies, being different from
the original bars-within-bars mechanism.}
fueling mechanism does not work efficiently (Shlosman \& Noguchi 1993).

\subsection{Minor-Merger Driven Gas Fueling}

Since most galaxies have satellite galaxies (Zaritsky et al. 1997
and references therein), it is likely that they have already experienced
some minor mergers during their lives (Ostriker \& Tremaine 1975; Tremaine 1981).
In fact, many lines of evidence for minor mergers have been obtained even in
ordinary-looking galaxies; e.g., ^^ ^^ X" structures (Mihos et al.
1995; see also for reviews, Schweizer 1990; Barnes 1996).
Since the disk gas is transferred toward the central region of galaxies
as the minor merger proceeds, minor mergers
are also responsible for gas fueling in disk galaxies 
(Hernquist 1989; Mihos \& Hernquist 1994; Hernquist \& Mihos 1995).
It is worthwhile noting that Seyfert galaxies possess a statistically
significant excess of faint ($M_v \geq -18$) 
companion (satellite) galaxies (Fuentes-Williams \& Stocke 1988).

Although the minor-merger driven fueling has no fatal problem,
it is generally hard to find unambiguous evidence for minor mergers in many cases
because the dynamical perturbation is less significant than
that of typical galaxy interactions with massive companion galaxies.
In order to complete a minor merger, it takes $\sim 10^9$ years.
This time scale may be long enough to smear relics of the minor merger.
Thus, the majority of advanced minor mergers may be observed as ordinary-looking
isolated galaxies (Walker, Mihos, \& Hernquist 1996). 
This makes it difficult to verify that minor mergers
are really responsible for triggering activity in most Seyfert nuclei.

\section{IS THE MINOR-MERGER SCENARIO CONSISTENT WITH OBSERVATIONS ?}

As summarized briefly in the previous section, there is little evidence that 
both the tidal interaction and the bar-driven fueling work in the majority of
Seyfert galaxies. In this section, we investigate
whether or not the minor-merger scenario appears consistent with 
all important observational properties of Seyfert galaxies.

\subsection{Morphology of Seyfert Hosts}

Morphological properties of Seyfert hosts have been studied since the 
pioneering work by Adams (1977) who suggested the Seyfert activity
is associated with nuclei of disk galaxies. Although disk properties of
Seyfert hosts  (e.g., scale lengths and color) are similar to
those of disk galaxies without an AGN (MacKenty 1990), 
Seyfert nuclei are more often
found in disk galaxies with ringed structures (e.g., inner and/or outer
rings; Heckman 1978; Simkin et al. 1980; Arsenault 1989; Moles et al. 1995),
and with amorphous disks (e.g., S0 galaxies; Simkin et al. 1980; MacKenty 1990).
The formation of ringed structures in disk galaxies is generally
interpreted in terms of so-called Lindblad resonances (e.g., Binney \& 
Tremaine 1987). Note, however, minor mergers are also responsible for the 
formation of ringed structures (see Figure 4 in Mihos \& Hernquist 1994).
Bar-mode instability in host disks may also be excited by minor mergers 
although the bar formation depends on both a relative mass and an 
an orbital parameter of the satellite (e.g., Byrd et al. 1986).
It is also known that minor mergers cause the kinematic heating
of host disks (e.g., Quinn, Hernquist, \& Fullagar 1993; Walker et al. 1996;
Vel\'aquez \& White 1999).
Such disk galaxies may be classified as S0 or amorphous galaxies
which are frequently observed in the Seyfert hosts.
Therefore, the minor merger scenario appears consistent with the
morphological variety of Seyfert hosts.

\subsection{The Formation of Randomly-Oriented Anisotropic Radiation}

Recent CCD narrow-band imaging studies of NLRs of Seyfert galaxies 
have shown that the NLRs often show bi-conical structures
(Pogge 1989; Wilson \& Tsvetanov 1994; Schmitt \& Kinney 1996).
The most spectacular property of the NLRs is that the bi-conical structures
are oriented independently from the host galactic disks in many cases. 
It is also known that the axes of nuclear radio jets in Seyfert nuclei 
show the same randomly-oriented nature (Schmitt et al. 1997).
These observational results suggest that accretion disks are rotating around
randomly-oriented axes which are different from the rotational axes of the
host disks (Clarke et al. 1998).

Generally, nuclear gas disks are not necessarily 
aligned to the host disks. For example, if some non-axisymmetric potential 
such as a tumbling bar is present in a disk, the preferred plane for the nuclear 
gas is perpendicular to the host disk and thus a tilted nuclear gas disk
can be made (Tohline \& Osterbrock 1982).
If this is the case, we would observe that NLRs tend to align 
along the bar axes. However, there is no correlation between the NLR axes and
the bar axes (Bower et al. 1995).
Note also that all Seyfert galaxies do not have such strong bars
(Simkin et al. 1980; Arsenault 1989; MacKenty 1990; Moles et al. 1995).
Therefore, it is unlikely that 
the NLR axes are controlled by the bar potential in the Seyfert galaxies.

Molecular-gas tori probed by the H$_2$O maser emission at 22 GHz
show the significant warping in some nearby AGNs such as NGC 1068
(Greenhill et al. 1996; see also Begelman, \& Bland-Hawthorn 1997) and NGC 4258
(Miyoshi et al. 1995). This warping can be explained by the radiative
effect from the central engine (Pringle 1996, 1997) given that
a typical size of tori is much less than 10 pc (e.g., Taniguchi \& Murayama 1998).
However, the warped disks traced by other molecular lines
such as CO and HCN are found to extend spatially
up to radius of $\sim$ 100 pc (Kohno et al. 1996; Tacconi 1998).
Such large-scale tilted gas disks cannot be formed by the effect of
radiation force from the central engine because a typical warping radius
explained by the radiation force 
is of the order of 0.01 pc for AGNs (Pringle 1997).

We discuss whether or not minor mergers are responsible for
the formation of tilted nuclear gas disks.
If a merging satellite galaxy has no nucleus (e.g., Magellanic clouds),
the gas in the satellite will interact with the gas in
the host disk and then be settled in the disk before reaching the nuclear region.
On the other hand, if it has a nucleus (e.g., M32),
the satellite nucleus will sink toward the nuclear region because of the
dynamical friction (Taniguchi \& Wada 1996). 
Here we regard that a nucleus is either a SMBH or 
a significant concentration of nuclear star cluster.
In this respect, satellite galaxies in Mihos \& Hernquist (1994)
and Hernquist \& Mihos (1995) are also nucleated ones.
Hence, we suggest that 
only minor mergers with {\it nucleated} satellites are responsible for
triggering activity in Seyfert nuclei.

The orbital decay of satellite galaxies could occur from random
orientations statistically. Even if a satellite takes a highly inclined orbit,
numerical studies generally show that the satellite orbit tends to settle
in the disk plane before it reaches the host center (e.g., Quinn et al. 1993).
However, note that the spatial resolution in the previous numerical studies
is about a few hundred pc. 
The bi-conical NLR of Seyfert galaxies may be 
collimated by a molecular/dusty torus around the central engine.
The plane in which the torus resides seems to be almost parallel
to the final orbital plane of the satellite nucleus around
the host nucleus. Since a typical inner radius of tori is of the 
order of 0.1 pc (Taniguchi \& Murayama 1998 and references therein), 
the spatial resolution in the previous numerical studies is too poor
to specify the final orbital plane in minor mergers.
If a host galaxy has no significant bulge component, 
the satellite nucleus may not be able to transfer its angular momentum
perpendicular to the disk efficiently. This suggests that the final 
orbital plane of the satellite nucleus around the host nucleus is
often different from the host disk. Therefore, 
the minor-merger scenario appears consistent with the observed random nature 
of the tilted nuclear gas disks in Seyfert galaxies;
see, for the formation of tilted nuclear gas disks, numerical simulations 
by Taniguchi \& Wada (1996).

\subsection{Type 1 vs. Type 2 Seyfert Nuclei}

The two types of Seyfert activity (S1 and S2) are
unified introducing optically-thick dusty/molecular tori around the
central engine (e.g., Antonucci 1993; Heisler, Lumsden, \& Bailey 1997).
However, there are some observational differences between S1s and S2s 
other than the presence/absence of broad-line regions.
The first important difference is that S1s tend to have
their torus axes aligned close to the host disk axes; i.e., 
the random nature is more frequently observed in S2s (Schmitt
\& Kinney 1996; see also Maiolino \& Rieke 1995). This difference may be 
explained in terms of the 
difference in orbits of satellite galaxies which merged into the hosts;
i.e., S1s prefer minor mergers with satellites whose orbits are relatively
parallel to the host disks while S2s prefer those with polar-like orbits.
Here we should remember that the current unified model explains the 
distinction between S1s and S2s as S1s (S2s) are observed from favored
(unfavored)  viewing angles. Therefore, even if some disk galaxies 
experience a minor merger with an polar-like orbit and then evolve into
Seyfert galaxies, some of them are observed as S1s if observed from
favored viewing angles. It is reminded that this causes some ambiguity 
in the explanation proposed above.

Next we discuss another interesting difference between S1s and S2s; 
S2s tend to experience circumnuclear starbursts (hereafter CNSBs) more 
frequently than S1s. 
Pogge (1989) made a narrow-band emission-line imaging survey of 20 nearby
Seyfert galaxies and found that CNSBs are found 
in $\sim$ 30\% of the S2s while no CNSB is found in the S1s.
Later observational studies (Oliva et al. 1995; Hunt et al. 1997) have 
confirmed that there is little evidence for CNSBs in S1s.
There are two necessary conditions to initiate CNSBs; 1) the presence of
cold molecular gas in the circumnuclear region 
enough to form a large number of massive stars,
and 2) the presence of some physical mechanism to trigger CNSBs
An earlier CO study of Seyfert galaxies suggested that 
S2s tend to be richer in CO than S1s (Heckman et al. 1989).
However, recent CO studies showed that there is little difference in
the molecular gas content between S1s and S2s (Maiolino et al. 1997;
Vila-Vilar\'o, Taniguchi, \& Nakai 1998).
Most disk galaxies including S0s may have molecular gas clouds with masses of 
$\sim 10^8 M_\odot$  in their circumnuclear regions (e.g., Taniguchi et al. 1994).
If this is the case, the most important factor for
the occurrence of CNSBs seems to be the triggering rather than
the gas content in the circumnuclear region of host disks.
Taniguchi \& Wada (1996) showed from their numerical simulations that 
the dynamical action of a pair of galactic nuclei (i.e., the host nucleus and 
the satellite one) can trigger CNSBs in the central region of minor mergers
(see also Taniguchi, Wada, \& Murayama 1997; section 4 in Taniguchi 1997).
Since it is known that the star formation activity in galactic disks
may be controlled by the surface mass density of cold gas (e.g., Kennicutt 1998;
Taniguchi \& Ohyama 1998),
more sophisticated observations will be necessary to unveil the difference
in surface gas mass density between S1s and S2s; e.g., sensitive, molecular-line
mapping surveys with radio interferometer facilities.

Finally we would like to also note that gaseous content in Seyfert hosts also affects
the visibility of the central engine if the circumnuclear gas disk is opaque
enough to hide the central engine (Maiolino \& Rieke 1995; Iwasawa et al. 1995;
Malkan, Varoujan, \& Raymond 1998).
Furthermore, the observational differences described above have been
considered as a serious problem for 
the strict unified model of Seyfert nuclei (e.g., Heckman et al. 1989).
They seem too complex to be understood unambiguously at present.
Perhaps, the complexity may be attributed to a wide variety of properties
of both host galaxies and  satellites as well as satellite orbits.
Detailed numerical simulations of minor mergers will be also recommended
for various sets of parameters.

\subsection{Environmental Properties of Seyfert Galaxies}

The observed excess of faint companion galaxies in the Seyfert galaxies  
(Fuentes-Williams \& Stocke 1988) provides
supporting evidence for the minor merger scenario because 
galaxies with more satellites should have more chances to have Seyfert nuclei
on a statistical ground.

Next we discuss galaxy-interaction-induced minor mergers.
Let us consider an interaction between disk galaxies. Each galaxy has
some satellites orbiting around the galaxy
under the given gravitational potential before the interaction.
After the two galaxies begin interacting with each other,
orbital motions of some satellites are disturbed  
and then forced suddenly to fall into the galaxies although 
some satellites are dynamically scattered from each host.
It is also expected that some satellites are directly trapped in the central 
regions of galaxies during the passage.
Therefore, galaxy interactions may enhance the chance of minor mergers.

One remaining problem is that there is a possible environmental difference
between S1s and S2s; S2s tend to have more massive companions than S1s
(e.g., Simkin 1991; Taniguchi 1992; Dultzin-Hacyan et al. 1999).
There is also a tendency that S2s prefer denser galaxy environments
(Laurikainen \& Solo 1995; De Robertis et al. 1998b).
If more massive galaxies tend to have more numerous, nucleated 
satellites, they could have more chances to evolve to Seyfert 
galaxies from a statistical view point. However, there seems no 
definite reason why S2s prefer such environs.
This problem will be open in future. 

\subsection{Frequency of the Seyfert Activity}

We estimate the frequency of the Seyfert activity 
if all the Seyfert activity is triggered by minor mergers with
nucleated satellite galaxies (see also Taniguchi \& Wada 1996).
Tremaine (1981) estimated that every galaxy would experience minor mergers with
its satellite galaxies several times. Since a typical galaxy may have several
satellite galaxies (Zaritsky et al. 1997), the probability of merger
for a satellite galaxy may be estimated to be $f_{\rm merger} \simeq 0.5$; i.e.,
half of the satellite galaxies have already merged to a host galaxy, while the rest
are still orbiting. Another important value is the number of nucleated
satellite galaxies. For example, 
M31 has two nucleated satellites (M32 and NGC 205), and a field S0 galaxy
NGC 3115 has a nucleated dwarf (van den Bergh 1986).
Although there has been no systematic search for {\it nucleated} satellite galaxies,
it is likely that  every galaxy has (or had) a few nucleated satellites:
we assume $n_{\rm sat} = 2$. If we assume that the typical lifetime of the Seyfert 
activity is $\tau_{\rm Seyfert} \simeq 10^8$ yr, we obtain the probability of the
Seyfert activity driven by minor mergers, 
$P_{\rm Seyfert} \simeq f_{\rm merger} ~ n_{\rm sat} ~ \tau_{\rm Seyfert} ~ 
\tau_{\rm Hubble}^{-1} \sim 0.01 (\tau_{\rm Seyfert}/10^8 ~ {\rm y})$,
where $\tau_{\rm Hubble}$ is the Hubble time, $\sim 10^{10}$ yr. Hence,
if minor mergers with nucleated satellites are responsible for
triggering the Seyfert activity, it is statistically expected that Seyfert nuclei 
are found in about 1 \% of field disk galaxies, being 
almost consistent with the observed value (e.g., Osterbrock 1989).

\subsection{Seyferts vs. Quasars}

It has often been considered
that the merger scenario is also applicable to the more luminous
starburst-AGN (i.e., ULIG-quasar) connection; i.e.,
major mergers between or among nucleated gas-rich galaxies
are progenitors of quasars (Sanders et al. 1988; Taniguchi \& Shioya 1998;
Taniguchi, Ikeuchi, \& Shioya 1999).
Optically bright quasars found in the local universe (e.g., $z < 0.2$)
show evidence for major mergers between/among galaxies (Hutchings \& Campbell 1983;
Heckman et al. 1986; Bahcall et al. 1997).
Although some quasar hosts look like giant elliptical
galaxies with little morphological peculiarity (Disney et al. 1995;
Bahcall et al. 1997), elliptical galaxies
may form from major mergers between/among disk galaxies (Barnes 1989;
Wright et al. 1990; Hernquist \& Barnes 1991; Ebisuzaki, Makino, \& Okumura 1991; 
Kormendy \& Sanders 1992; Weil \& Hernquist 1996).
Therefore, if we adopt that elliptical galaxies
hosting quasars were also made by major mergers, we could conclude that all the nearby
quasars were made by major mergers.

\subsection{A Summary of the Proposed Scenario}

In Fig. 1, we show our unified formation mechanism of AGNs proposed here
[see for a unified formation mechanism for both circumnuclear/nuclear starbursts
and ultraluminous starbursts in ULIGs (Taniguchi et al. 1997)].
Note that the viewing angle dependence is not explicitly introduced 
in Fig. 1 although this is also an important factor in our scenario.
If we postulate that the nuclear activity in all the Seyfert galaxies are 
triggered by minor mergers with nucleated satellites,
we have a possibility that various observational  properties of the Seyfert 
galaxies can be explained without invoking other physical mechanisms; e.g.,
the bar-driven or the tidal-interaction driven fueling.
Furthermore, it seems possible that all the nearby quasars come from major mergers 
between/among nucleated galaxies (Sanders et al. 1988; Taniguchi \& Shioya 1998;
Taniguchi, Ikeuchi, \& Shioya 1999). Therefore, if we adopt an idea that all the AGNs 
in the local universe arise from either minor or major mergers, 
we will have a unified (or single) formation mechanism 
of AGNs observed in the local universe.

Recently, De Robertis et al. (1998b) gave their careful thought on the 
interaction hypothesis; i.e., {\it there is a causal link between  activity
in the nucleus of a galaxy containing a supermassive compact object and 
disturbances to the host galaxy resulting from tidal interactions or
mergers} (see also De Robertis et al. 1998a). They discussed the potential
importance of minor mergers for triggering activity in Seyfert  nuclei
because a significant fraction of Seyfert hosts show little or no evidence for
a recent (major) merger. However, they also discussed a possibility that 
there are various triggering mechanisms depending on the luminosity of 
the class from LINERs to quasars. Although our scenario proposed here has an opposite 
sense from their idea, it is one of options of the interaction hypothesis.
Therefore, appreciating their careful thought, we would like to call our 
model ^^ ^^ {\it the simple interaction (or merger)  model}".
Finally we mention that any fueling mechanism is required to  
ensure that the gas in the host disk is surely
fueled into the very inner region (e.g., $\ll$ 1 pc).

\vspace{0.5cm}

I would like to thank my colleagues, in particular,
Satoru Ikeuchi,  Keiichi Wada, Toru Yamada, Yasuhiro Shioya, 
and Takashi Murayama for useful discussion.
I would also like to thank Dave Sanders and Chris Mihos for useful
discussion about the merger-driven gas fueling and Josh Barnes and an 
anonymous referee for useful comments and suggestions.
This work was partly done at C\'ordoba Observatory in Argentina.
I would  like to thank Silvia Fern\'andes and Sebastian L\'ipari
for their warm hospitality.
This work was financially supported in part by Grant-in-Aids for the Scientific
Research (Nos. 10044052, and 10304013) of the Japanese Ministry of
Education, Culture, Sports, and Science.


\newpage


\begin{figure}
\epsfysize=18.5cm \epsfbox{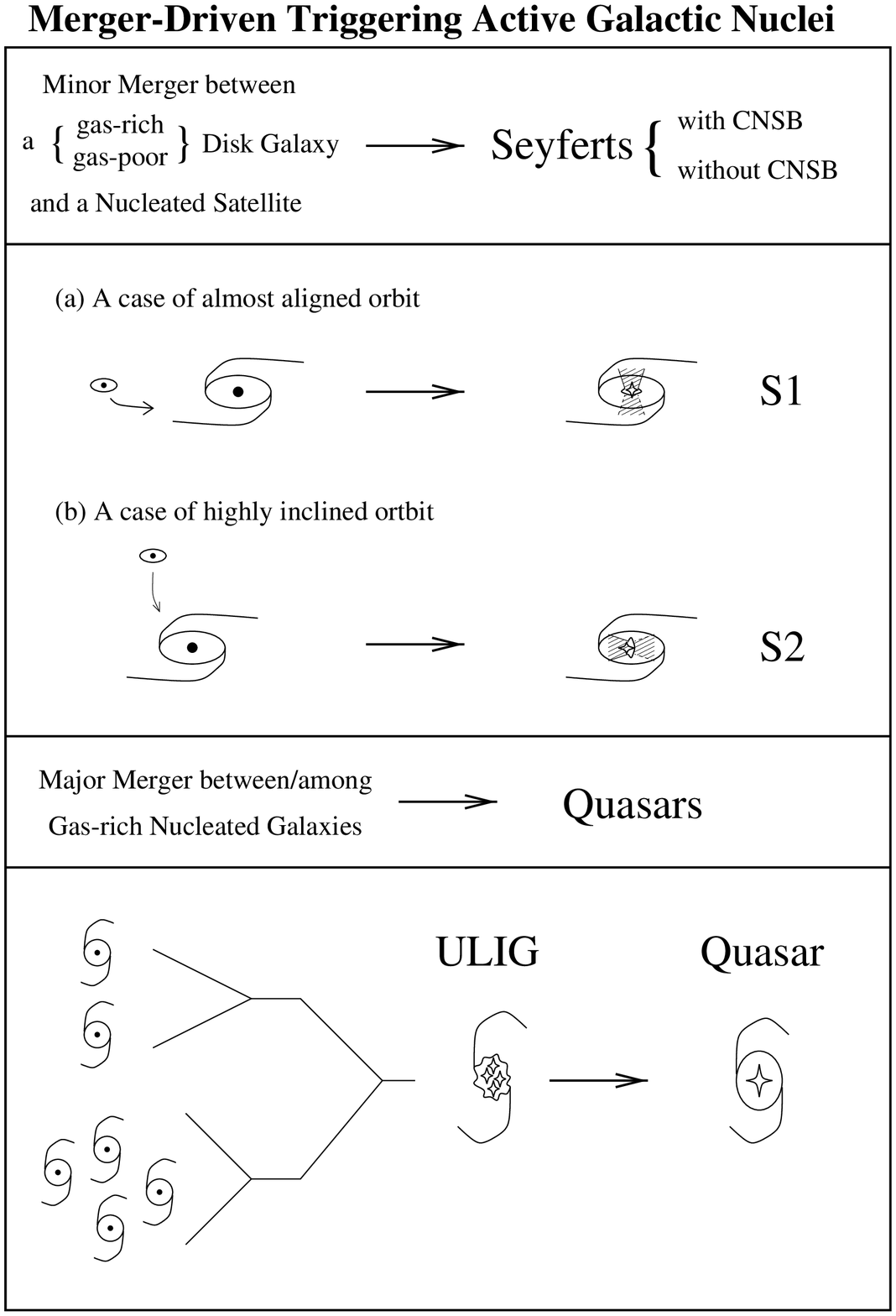}
\caption[]{
The merger-driven unified scenario for triggering AGNs.
Note that the viewing angle dependence is not explicitly introduced
in Fig. 1 although this is also an important factor in our scenario.
Note also that CNSB = circumnuclear starburst.
\label{fig1}
}
\end{figure}


\begin{references}
\reference{1}{Adams, T. F. 1977, ApJS, 33, 19}
\reference{1}{Antonucci, R.\ R.\ J. 1993, ARA\&A, 31, 473}
\reference{1}{Arsenault, R. 1989, A \& A, 217, 66}
\reference{1}{Bahcall, J. N., Kirhakos, S., Saxe, D. H., \& Schneider, D. P.
              1997, ApJ, 479, 642}
\reference{1}{Barnes, J. E. 1989, Nature, 338, 123}
\reference{1}{Barnes, J. E. 1996, in Formation of the Galactic Halo, ASP Conf. Ser.,
              102, edited by H. Morrison, and A. Sarajedini, 415}
\reference{1}{Barnes, J. E., \& Hernquist, L. E. 1992, ARA \& A, 30, 705}
\reference{1}{Begelman, M. C., \& Bland-Hawthorn, J. 1997, Nature, 385, 22}
\reference{1}{Bergvall, N., \& Johansson, L. 1995, A\&AS, 113, 499}
\reference{1}{Binney,  J., \& Tremaine, S. 1987, Galactic Dynamics (Princeton University
              Press: Princeton)}
\reference{1}{Bower, G., Wilson, A. S., Morse, J. A. Gelderman, R., Whittle, M.,
              Mulchaey, J. 1995, ApJ, 454, 106}
\reference{1}{Bushouse, H. A. 1986, AJ, 91, 255}
\reference{1}{Bushouse, H. A. 1987, ApJ, 320, 49}
\reference{1}{Byrd, G. G., Saarinen, S., \& Valtonen, M. J. 1986, MNRAS, 220, 619}
\reference{1}{Clarke, C. J., Kinney, A. L., \& Pringle, J. E. 1998, ApJ, 495, 189}
\reference{1}{Cutri, R. M., McAlary, C. W. 1985, ApJ, 296, 90}
\reference{1}{Dahari, O. 1984, AJ, 89, 966}
\reference{1}{Dahari, O. 1985, ApJS, 57, 643}
\reference{1}{De Robertis, M. M., Hayhoe, K., \&  Yee, H. K. C. 1998a, ApJS, 115, 163}
\reference{1}{De Robertis, M. M., Yee, H. K. C., \& Hayhoe, K. 1998b, ApJ, 496, 93}
\reference{1}{Disney, M., et al. 1995, Nature, 376, 150}
\reference{1}{Dultzin-Hacyan, D., Krongold, Y., Fuentes-Guridi, I., \& Marziani, P.
              1999, ApJ, 513, L111}
\reference{1}{Ebisuzaki, T., Makino, J., \& Okumura, S. K. 1991, Nature, 354, 212}
\reference{1}{Friedli, D. 1996, Barred Galaxies, ASP Conference Ser., 91, edited by
              R. Buta, D. A. Crocker, and B. G. Elmegreen, 378}
\reference{1}{Fuentes-Williams, T., \& Stocke, J. T. 1988, AJ, 96, 1235}
\reference{1}{Gaskell, C. M. 1985, Nature, 315, 386}
\reference{1}{Greenhill, L. J., Gwinn, C. R., Antonucci, R, \& Barvanis, R.
              1996, ApJ, 472, L21}
\reference{1}{Heckman, T. M. 1978, PASP, 90, 241}
\reference{1}{Heckman, T. M. 1980, A \& A, 87, 152}
\reference{1}{Heckman, T. M., Blitz, L., Wilson, A. S., Armus, L., \& Miley, L.
              1989, ApJ, 342, 735}
\reference{1}{Heckman, T. M., Smith, E. P., Baum, S. A., van Bruegel, W. J. M.,
              Illingworth, G. D., Bothun, G. D., \& Balick, B. 1986, ApJ, 311, 526}
\reference{1}{Heisler, C. A., Lumsden, S. L., \& Bailey, J. A. 1997, Nature, 385, 700}
\reference{1}{Hernquist, L. 1989, Nature, 340, 687}
\reference{1}{Hernquist, L. E., \& Barnes, J. E. 1991, Nature, 354, 210}
\reference{1}{Hernquist, L., \& Mihos, C. J. 1995, ApJ, 448, 41}
\reference{1}{Ho, L. C., Filippenko, A. V., \& Sargent, W. L. W. 1997, ApJ, 487, 591}
\reference{1}{Hunt, L. K., Malkan, M. A., Salvati, M., Mandolesi, N., Palazzi, E.,
              \& Wade, R. 1997, ApJS, 108, 229}
\reference{1}{Hunt, L. K., Malkan, M. A., Moriondo, G., \& Salvati, M.
              1999, ApJ, 510, 637}
\reference{1}{Hutchings, J. B., \& Campbell, B. 1983, Nature, 303, 584}
\reference{1}{Iwasawa, K., Kunieda, H., Tawara, Y., Awaki, H., Koyama, K., Murayama, T.,
              \& Taniguchi, Y. 1995, AJ, 110, 551}
\reference{1}{Keel, W. C. 1996, AJ, 111, 696}
\reference{1}{Keel, W. C., Kennicutt, R. C. Jr., Hummel, E., \& van der Hulst, J. M. 
              1985, AJ, 90, 708}
\reference{1}{Kennicutt, R. C. Jr. 1998, ApJ, 498, 541}
\reference{1}{Kennicutt, R. C. Jr., \& Keel, W. C. 1984, ApJ, 279, L5}
\reference{1}{Kohno, K., Kawabe, R., Tosaki, T., \& Okumura, S. K. 1996, ApJ, 461, L29}
\reference{1}{Kormendy, J., Bender, R., Evans, A. S., \& Richstone, D. 1998,
              AJ, 115, 1823}
\reference{1}{Kormendy, J., \& Sanders, D. B. 1992, ApJ, 390, L53}
\reference{1}{Laurikainen, E., \& Salo, H. 1995, A \& A, 293, 683}
\reference{1}{Laurikainen, E., Salo, H., Teerikorpi, P., \& Petrov, G. 1994, A \& AS,
              108, 491}
\reference{1}{MacKenty, J. W. 1989, ApJ, 343, 125}
\reference{1}{MacKenty, J. W. 1990, ApJS, 72, 231}
\reference{1}{Maiolino, R., \& Rieke, G. H. 1995, ApJ, 454, 95}
\reference{1}{Maiolino, R., \& Risailti, G., \& Salvati, M. 1998, A\&A, 341, L35}
\reference{1}{Maiolino, R., Ruiz, M., Rieke, G. H., \& Papadopoulos, P. 1997, 
              ApJ, 485, 552} 
\reference{1}{Malkan, M. A., Varoujan, G., \& Raymond, T. 1998, ApJS, 117, 25}
\reference{1}{Mihos, C. J., \& Hernquist, L. 1994, ApJ, 425, L13}
\reference{1}{Mihos, C. J., Walker, I. R.,  Hernquist, L., Mendes-Oliveira, C. 
              \& Bolte, M. 1995, ApJ, 447, L87}
\reference{1}{Miyoshi, M., Moran, J., Herrnstein, J., Greenhill, L., Nakai, N.,
              Diamond, P., \& Inoue, M. 1995,  Nat, 373, 127}
\reference{1}{Moles, M., M\'arquez, I., \& P\'erez, E. 1995, ApJ, 438, 604}
\reference{1}{Mulchaey, J. S., \& Regan, M. 1997, ApJ, 482, L135}
\reference{1}{Noguchi, M. 1988, A \& A, 203, 259}
\reference{1}{Norman, C. A. 1990, in Massive Stars in Starbursts, edited by C. Leitherer,
              N. R. Walborn, T. M. Heckman, \& C. A. Norman (Cambridge University Press:
              Cambridge, UK), 271}
\reference{1}{Oliva, E., Origlia, L., Kotilainen, J. K., \& Moorwood, A. F. M.
              1995, A \& A, 301, 55}
\reference{1}{Osterbrock, D.\ E. 1989, {\it Astrophysics of Gaseous Nebulae and
              Active Galactic Nuclei} (San Francisco, Freeman), 310}
\reference{1}{Ostriker, J. P., \& Tremaine, S. 1975, ApJ, 202, L113}
\reference{1}{Petrosyan, A. R. 1982, Astrophysics, 18, 312}
\reference{1}{Pogge, R. W. 1989, ApJ, 345, 730}
\reference{1}{Pringle J.E. 1996, MNRAS 281, 357}
\reference{1}{Pringle J.E. 1997, MNRAS 292, 136}
\reference{1}{Quinn, P. J., Hernquist, L., \& Fullagar, D. P. 1993, ApJ, 403, 74}
\reference{1}{Rafanelli, P., Violato, M., \& Baruffolo, A. 1995,
              AJ, 109, 1546}
\reference{1}{Rees, M. J. 1984, ARA \& A,  22, 471}
\reference{1}{Roos, N. 1981, A \& A, 104, 218}
\reference{1}{Roos, N. 1985a, ApJ, 294, 479}
\reference{1}{Roos, N. 1985b, ApJ, 294, 486}
\reference{1}{Sanders, D. B., et al. 1988, ApJ, 325, 74}
\reference{1}{Schmitt, H. R., \& Kinney, A. L. 1996, ApJ, 463, 498}
\reference{1}{Schmitt, H. R., Kinney, A. L., Storchi-Bergmann, T., \& Antonucci, R.
              1997, ApJ, 477, 623}
\reference{1}{Schwarz, M. P. 1981, ApJ, 247, 77}
\reference{1}{Schweizer, F. 1990, in Dynamics and Interaction of Galaxies, ed.
              R. Wielen (Heidelberg: Springer), 60}
\reference{1}{Sekiguchi, K., \& Wolstencroft, R. D. 1992, MNRAS, 255, 581}
\reference{1}{Shlosman, I., Begelman, M. C., \& Frank, J. 1990, Nature, 345, 679}
\reference{1}{Shlosman, I., Frank, J., \& Begelman, M. C. 1989, Nature, 338, 45}
\reference{1}{Shlosman, I., \& Noguchi, M. 1993, ApJ, 414, 474}
\reference{1}{Simkin, S. M. 1991, in Paired and Interacting Galaxies,
              ed. J. W. Sulentic, and W. C. Keel (NASA), 399}
\reference{1}{Simkin, S. M., Su, H. J., \& Schwarz, M. P. 1980, ApJ, 237, 404}
\reference{1}{Smith, E. P., \& Hintzen, P. 1991, AJ, 101, 410}
\reference{1}{Tacconi, L. 1998, private communication}
\reference{1}{Taniguchi, Y. 1992, in Relationships between Active Galactic Nuclei
              and Starburst Galaxies, ed. A. V. Filippenko,
              ASP Conference Ser., 31, 397}
\reference{1}{Taniguchi, Y. 1997, ApJ, 487, L17}
\reference{1}{Taniguchi, Y., Ikeuchi, S., \& Shioya, Y. 1999, ApJ, 514, L9}
\reference{1}{Taniguchi, Y., \& Murayama, T. 1998, ApJ, 501, L25}
\reference{1}{Taniguchi, Y., Murayama, T., Nakai, N., Suzuki, M., \& Kamaya, O. 
              1994, AJ, 108, 468}
\reference{1}{Taniguchi, Y., \& Ohyama, Y. 1998, ApJ, 509, L89}
\reference{1}{Taniguchi, Y., \& Shioya, Y. 1998, ApJ, 501, L167}
\reference{1}{Taniguchi, Y., \& Wada, K. 1996, ApJ, 469, 581}
\reference{1}{Taniguchi, Y., Wada, K., \& Murayama, T. 1997, RMxAC, 6, 240}
\reference{1}{Telesco, C. M., Wolstencroft, R. D., \& Done, C. 1988, ApJ, 329, 174}
\reference{1}{Tohline, J. E., \&  Osterbrock, D. E. 1982, ApJ, 252, L49}
\reference{1}{Tremaine, S. 1981, in The Structure and Evolution of Normal Galaxies,
              ed. S. M. Fall, \& D. Lynden-Bell (Cambridge: Cambridge University 
              Press), 67}
\reference{1}{Vel\'aquez, H., \& White, S. D. M. 1999, MNRAS, 304, 254}
\reference{1}{van den Bergh, S. 1986, AJ, 91, 271}
\reference{1}{Vila-Vilar\'o, B., Taniguchi, Y., \& Nakai, N. 1998, AJ, 116, 1553}
\reference{1}{Wada, K., \& Habe, A. 1992, MNRAS, 258, 82}
\reference{1}{Wada, K., \& Habe, A. 1995, MNRAS, 277, 433}
\reference{1}{Walker, I. R., Mihos, C. J., \& Hernquist, L. 1996, ApJ, 460, 121}
\reference{1}{Weil, M. L., \& Hernquist, L. 1996, ApJ, 460, 101}
\reference{1}{Wilson, A. S., \& Tsvetanov, Z. I. 1994, AJ, 107, 1227}
\reference{1}{Wright, G. S., James, P. A., Joseph, R. D., \& McLean, I. S.
              1990, Nature, 344, 417}
\reference{1}{Zaritsky, D., Smith, R., Frenk, C., \& White, S. D. M. 1997,
              ApJ, 478, 39}
\end{references}
\end{document}